\documentclass[twocolumn,prl,amsmath,amssymb]{revtex4}

\usepackage{graphicx} 

\begin{document}

\newcommand{\ra}{\rangle}
\newcommand{\la}{\langle}
\newcommand{\bp}{{\bf p}}
\newcommand{\bq}{{\bf q}}
\newcommand{\hx}{{\hat {\bf x}}}
\newcommand{\hy}{{\hat {\bf y}}}
\newcommand{\hz}{{\hat {\bf z}}}
\newcommand{\hp}{{\hat {\bf p}}}
\newcommand{\hq}{{\hat {\bf q}}}
\newcommand{\no}{\noindent}
\newcommand{\non}{\nonumber}
\newcommand{\hi}{\hangindent=45pt}
\newcommand{\me}{\mathrm{e}}
\newcommand{\mi}{\mathrm{i}}

\title{Entangled Light in Moving Frames}
\author{Attila J. Bergou}
\author{Robert M. Gingrich}
\author{Christoph Adami}

\affiliation{
    Quantum Computing Technologies Group, Jet Propulsion Laboratory
    126-347\\ California Institute of Technology, Pasadena,
    CA 91109-8099}

\date{\today}

\begin{abstract}
We calculate the entanglement between a pair of polarization-entangled photon beams
as a function of the reference frame, in a fully relativistic framework. We find
the transformation law for helicity basis states and show that, while it is
frequency independent, a Lorentz transformation on a momentum-helicity eigenstate
produces a momentum-dependent phase. This phase leads to changes in the reduced
polarization density matrix, such that entanglement is either decreased or increased,
depending on the boost direction, the rapidity, and the spread of the beam.
\end{abstract}

\pacs{PACS numbers: 03.30+p,03.67-a,03.65.Ud}

\maketitle

The second quantum revolution~\cite{dowling03} is changing the ways in which we
think about quantum systems. Rather than just describing and predicting their
behavior, we now use new tools such as quantum information theory to organize and
control quantum systems, and turn their non-classical features to our advantage in
creating {\it quantum technology}. The central feature that makes quantum
technology possible is quantum {\em entanglement}, which implies that particles or
fields that have once interacted are connected by an overall wave function even if
they are detected arbitrarily far away from each other. Such {\em entangled pairs},
first discussed after their introduction by Einstein, Podolsky, and
Rosen~\cite{epr35}, are crucial in technology such as quantum
teleportation~\cite{Bennettetal93} and superdense coding~\cite{BennettWiesner92}.
Furthermore, quantum entanglement is critical in applications such as quantum
optical interferometry, where quantum entangled $N$-photon pairs can increase the
shot-noise limited sensitivity up to the Heisenberg limit~\cite{dowling98}.

While quantum entanglement as a resource has been studied extensively within the
last decade~\cite{NielsenChuang00}, it was realized only recently that this
resource is frame-dependent, and changes non-trivially under Lorentz
transformations~\cite{Czachor97,Peresetal02,AlsingMilburn02,TerashimaUeda02,GingrichAdami02,PeresTerno02}.
In particular, Gingrich and Adami showed that the entanglement between the spins of
a pair of massive spin-1/2 particles depends on the reference frame, and can either
decrease or increase depending on the wave-function of the
pair~\cite{GingrichAdami02}. A consequence of this finding is that the entanglement
resource could be manipulated by applying frame changes only.  Many applications of
quantum technology, however, involve entangled photons rather than massive spin-1/2
particles, to which the massive theory does not apply. In this letter, we work out
the consequences of Lorentz transformations on photon beams that are entangled in
polarization. Each photon beam is described by a Gaussian wave packet with a
particular angular spread in momentum, and for the sake of being definite we
discuss a state whose polarization entanglement can be thought of as being produced
by down-conversion. Because both spin-1/2 particles and photons can be used as
quantum information carriers (qubits), the present calculation also contributes to
the nascent field of Relativistic Quantum Information Theory~\cite{PeresTerno03}.

In order to calculate how a polarization-entangled photon state transforms under
Lorentz transformations, we need to discuss the behavior of the photon basis sates.
Because there is no rest frame for a massless particle, the analysis of the spin
(polarization) properties is quite distinct from the massive case. For instance,
instead of using $p^{\mu} = (m,{\bf 0})$ as the standard 4-vector
(see~\cite{GingrichAdami02}), we have to define the massless analog $k^{\mu} =
(1,\hz)$. Note that $k^{\mu}$ has no parameter $m$ and is no longer invariant under
all rotations. In fact, the {\it little group} of $k^{\mu}$ is isomorphic to the
non-compact two-dimensional Euclidean group $E(2)$ (the set of transformations that
map a two-dimensional Euclidean plane onto itself) . For a massless spin-one
particle the standard vector allows us to define the eigenstate
\begin{eqnarray}
    P^\mu |\hz \lambda \ra      & = & k^\mu|\hz \lambda \ra  \\
    J_z |\hz \lambda \ra        & = & \lambda|\hz \lambda \ra \;,
\end{eqnarray}
where $\hz$ is a unit vector pointing in the z-direction. Since the particle is
massless, $\lambda$ is restricted to $\pm 1$ \cite{wigner}.

The momentum-helicity eigenstates are defined as
\begin{equation}\label{eq:momentumdef}
    |{\bf p} \lambda \ra = H({\bf p}) |\hz \lambda \ra \;,
\end{equation}
where $H(\bp)$ is a Lorentz transformation that takes $\hz$ to $\bp$.  The choice
of $H(\bp)$ is not unique, and different choices lead to different interpretations
of the parameter $\lambda$.  For instance, in the massive case the choice of
$H(\bp)$ can lead to $\lambda$ being either the rest frame spin or the helicity. In
the present case it is convenient to choose
\begin{equation}
    H(\bp) = R(\hp) L_z (\xi_{\bp}),
\end{equation}
where $L_z (\xi_{\bp})$ is a Lorentz boost along $\hz$ that takes $\hz$ to $|\bp|
\hz$ and $R(\bp)$ is a rotation that takes $\hz$ to $\hp$, while $\xi_{\bp}$ is the
rapidity of the moving frame,
\begin{equation}
    \xi_{\bp}= \ln |\bp|\; .
\end{equation}
For a parameterization in polar coordinates, we can write $\hp = (\sin \theta \cos
\phi,\sin \theta \sin \phi, \cos \theta)$
\begin{equation}
    \label{Rdef}
    R(\hp) \equiv R_z (\phi) R_y (\theta).
\end{equation}
Again, this choice of $R(\hp)$ is not unique (see for example \cite{yndurain}) but
particularly easy to deal with in this context. An arbitrary two-particle state in
this formalism can be written as
\begin{equation}\label{eq:state}
    |\Psi_{AA'BB'} \ra = \int \! \! \! \int \sum_{\lambda \sigma}
    g_{\lambda \sigma}(\mathbf{p},\mathbf{q})|\mathbf{p} \lambda
    \ra_{AA'} |\mathbf{q} \sigma \ra_{BB'} \tilde{\rm d} \mathbf{p}
    \tilde{\rm d} \mathbf{q}\ ,
\end{equation}
where $|\mathbf{p} \lambda \ra_{AA'}$ and $|\mathbf{q} \sigma \ra_{BB'}$ correspond
to the momentum and helicity states, as defined in Eq.~(\ref{eq:momentumdef}), of
photons  $A$ and $B$.  Furthermore, $\tilde{d}\mathbf{p}$ and $\tilde{d}\mathbf{q}$
are the Lorentz-invariant momentum integration measures:
\begin{equation}
    \widetilde{\rm d}\bp \equiv \frac{{\rm d}^3 \bp}{2 |\bp|}
\end{equation}
and the functions $g_{\lambda \sigma}(\bp,\bq)$ must satisfy
\begin{equation}
    \int \! \! \! \int \sum_{\lambda \sigma} |g_{\lambda \sigma}(\bp,\bq)|^2 \,
    \widetilde{\rm d}\bp\, \widetilde{\rm d}\bq = 1\, .
\end{equation}

To work out how a Lorentz boost affects an entangled state, we must understand how
the basis states $|\bp \lambda \ra$ transform.  Following \cite{wigner,tung}, we
apply a boost $\Lambda$ to $|\bp \lambda \ra$
\begin{equation}
    \Lambda |\bp \lambda \ra = H(\Lambda \bp) H(\Lambda \bp)^{-1} \Lambda
    H(\bp) |\hz \lambda \ra \;,
\end{equation}
where $H(\Lambda \bp)^{-1} \Lambda H(\bp)$ is a member of the {\em little group} of
$\hz$ (leaves $\hz$ invariant), and hence is a rotation and/or translation in the
$x$,$y$ plane. The translations can be shown~\cite{wigner} not to affect the
spin/helicity, and we are thus left with just a rotation by an angle $\Theta
(\Lambda,\bp)$.  Using the parameterization $\bp = p (\sin \theta \cos \phi,\sin
\theta \sin \phi, \cos \theta)$ and solving for $\Theta (\Lambda,\bp)$ we obtain
\begin{eqnarray}\label{thetadef}
    \Theta (\Lambda, \bp) = \left\{
    \begin{array}{r@{\quad:\quad}l}
    0      & \Lambda = L_z (\xi) \\
    0      & \Lambda = R_z (\gamma), \hp \not= \hz \\
    \gamma & \Lambda = R_z (\gamma), \hp = \hz \\
    \tan^{-1} \left(\frac{A}{B} \right)
    & \Lambda = R_y (\gamma)
    \end{array}
    \right.
\end{eqnarray}
for different Lorentz transformations and momenta, where
\begin{eqnarray}
    A & = & \sin \gamma \sin \theta \\
    B & = & \cos \theta \sin \gamma \cos \phi + \cos \gamma \sin \phi \; .
\end{eqnarray}
Noting that
\begin{equation}
    R_z(\Theta (\Lambda' \Lambda, \bp)) = R_z(\Theta(\Lambda',\Lambda \bp))
    R_z(\Theta(\Lambda, \bp))
\end{equation}
and taking advantage of the fact that all Lorentz boosts can be constructed using
$L_z$, $R_z$ and $R_y$, Eq.~(\ref{thetadef}) allows us to find
$\Theta(\Lambda,\bp)$ for any $\Lambda$, and any momentum $\bp$. Applying this
rotation to the momentum-helicity eigenstate of a massless particle we obtain
\begin{equation}
    \Lambda |\bp \lambda \ra = \me^{-\mi \lambda \Theta (\Lambda, \bp)}
    |\Lambda \bp \lambda \ra \, .
\end{equation}

Typically, it is the polarization of a photon that is measured in quantum optics
experiments, not the helicity. Let us therefore examine the effects of a Lorentz
transformation on a photon's polarization 4-vector.

The polarization 4-vectors for positive and negative helicity states are
given by
\begin{equation}
    \epsilon_{\pm}^{\mu}(\hp) = \frac{R(\hp)}{\sqrt{2}} \left[
    \begin{array}{c}
    0 \\ 1 \\ \pm i \\ 0
    \end{array}
    \right] \ .
\end{equation}
A general polarization vector is, of course, formed by the superposition of the two
basis vectors.  According to \cite{han,AlsingMilburn02}, for a given 4-momentum
$p^{\mu}$ and associated polarization $\epsilon^{\mu}$, a Lorentz boost has the
following effect:
\begin{equation}
    \label{hantransf}
    D(\Lambda) \epsilon^{\mu} = R(\Lambda \hp) R(\hp)^{-1}
    \epsilon^{\mu} \, .
\end{equation}
However, this transformation is only correct for pure boosts in the
z-direction, or  rotations around the z-axis if this axis is not the
momentum axis (as for those cases the angle $\Theta(\Lambda,\bp)$ in
Eq.~(\ref{thetadef}) vanishes). In general,  the four-vector
$\epsilon^{\mu}$ transforms as
\begin{equation}
    \label{epsilonrot}
    D(\Lambda) \epsilon^{\mu} = R(\Lambda \hp) R_z(\Theta
    (\Lambda,\bp)) R(\hp)^{-1} \epsilon^{\mu} \, .
\end{equation}
 It is helpful to write $D(\Lambda)$ in an alternative form
\begin{equation}\label{epsilongauge}
    D(\Lambda) \epsilon^{\mu} = \Lambda \epsilon^{\mu} - \frac{(\Lambda
    \epsilon^{\mu})^0}{(\Lambda p^{\mu})^0} \Lambda p^{\mu} \, ,
\end{equation}
where $(\Lambda \epsilon^{\mu})^0$ and $(\Lambda p^{\mu})^0$ denote the time-like
component of the transformed polarization and momentum 4-vectors, respectively. The
form Eq.~(\ref{epsilongauge}) agrees with the general law described in
\cite{weinberg}. The proof that Eqs.~(\ref{epsilonrot}) and (\ref{epsilongauge})
are equivalent is non-trivial, but an outline is as follows. Note that both forms
of $D(\Lambda)$ obey
\begin{equation}
    D(\Lambda') D(\Lambda) \epsilon^{\mu} = D(\Lambda' \Lambda)
    \epsilon^{\mu}
\end{equation}
and both forms have the property
\begin{equation}\label{ourtransf}
    D(R) \epsilon^{\mu} = R \epsilon^{\mu}
\end{equation}
where $R$ is a rotation.  An explicit calculation of $D(L_z(\xi))$ then shows that
they are equivalent.

The second term on the right hand side of Eq.~(\ref{epsilongauge}) is just a
momentum-dependent gauge transformation.  It must be different for each momentum in
order to keep a consistent overall (Coulomb) gauge.  To see that this term leads to
measurable consequences consider the polarization vector for classical
electromagnetic waves.  The polarization vector points along the {\it
  gauge-invariant} electric field, and the direction of this vector
undergoes the same transformation as in Eq.~(\ref{epsilongauge}) (or
\ref{epsilonrot}) when acted on by a Lorentz transformation.  In fact, the
magnitude of the electric field undergoes the same transformation as the diameter
of an infinitesimal circle centered at the momentum. This holds for any Lorentz
transformation and momentum. A detailed study of this transformation will be
published elsewhere.

\begin{figure}[htb]
        \includegraphics[width=\linewidth]{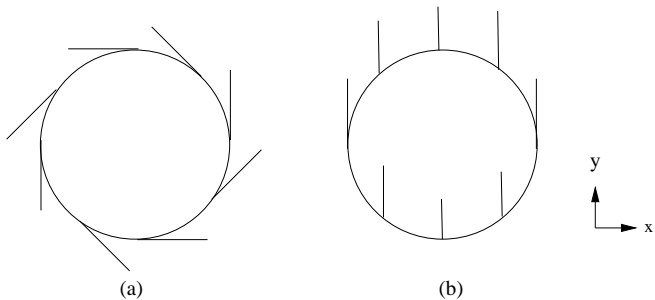}
        \caption{\label{fig:polarization} (a): ``Standard'' vertical polarization
        vectors $|v_{\bp}'\ra$ point in polar directions. (b): ``True" vertical polarization vectors
        $|v_{\bp}\ra$ remain mostly in the x-y plane.}
\end{figure}

In the following, we investigate two entangled photon beams moving along the
z-axis. The beams are in a momentum product state, and fully entangled in
polarization,
\begin{equation}\label{eq:dist1}
    g_{\lambda \sigma}(\mathbf{p},\mathbf{q}) = \frac{1}{\sqrt{2}}
    \delta_{\lambda \sigma}  \me^{\mi \lambda \phi_{\bp} } \me^{\mi
    \sigma \phi_{\bq} } f(\mathbf{p}) f(\mathbf{q}) .
\end{equation}
In Eq.~(\ref{eq:dist1}), $\phi_{\bp}$ and $\phi_{\bq}$ are the polar
angles of $\mathbf{p}$ and $\mathbf{q}$ respectively.  The phase factors $\me^{\mi
\lambda \phi_{\bp} } \me^{\mi \sigma \phi_{\bq} }$ allow us to write
the state as
\begin{equation}\label{eq:hvstate}
    |\Psi \ra = \int \! \! \! \int
    \frac{1}{\sqrt{2}}(| h_{\bp} \ra | h_{\bq} \ra - | v_{\bp} \ra |
    v_{\bq} \ra) f(\mathbf{p}) |\mathbf{p} \ra  f(\mathbf{q})
    |\mathbf{q} \ra \tilde{d} \mathbf{p} \tilde{d} \mathbf{q} \;,
\end{equation}
where $| h_{\bp} \ra$ and $| v_{\bp} \ra$ are approximations of
horizontal and vertical polarization given by~\cite{phinote}
\begin{eqnarray}\label{eq:polar}
    | h_{\bp} \ra \equiv \frac{1}{\sqrt{2}} ( \me^{\mi \phi_{\bp}}
    \epsilon^{\mu}_{+}(\hp) + \me^{-\mi \phi_{\bp}}
    \epsilon^{\mu}_{-}(\hp)) \\ \label{eq:polar2}
    | v_{\bp} \ra \equiv \frac{-\mi}{\sqrt{2}} ( \me^{\mi \phi_{\bp}}
    \epsilon^{\mu}_{+}(\hp) - \me^{-\mi \phi_{\bp}}
    \epsilon^{\mu}_{-}(\hp))\ .
\end{eqnarray}
So, for small $\theta$ (small azimuthal spread of the momentum distribution) we
have:
\begin{eqnarray}
    | h_{\bp} \ra \simeq \hx\;, \\
    | v_{\bp} \ra \simeq \hy\;,
\end{eqnarray}
and Eq.~(\ref{eq:hvstate}) is a close approximation to a polarization Bell state.
Omitting the phase factors in (\ref{eq:dist1}) and (\ref{eq:polar}) instead
describes a photon beam where horizontal and vertical polarizations point in the
${\hat r}$ and ${\hat \phi}$ directions, respectively (see
Fig.~\ref{fig:polarization}).

We specifically consider the beams to have a Gaussian spread in the azimuthal
direction,
\begin{equation}\label{eq:dist2}
    f(\mathbf{p}) = \frac{1}{N(\sigma)}
    {\rm exp}\left(-\frac12\left(\frac{\theta}{\sigma_\theta}\right)^2\right) \delta (|\bp| - p_{0})\;,
\end{equation}
where $\sigma_\theta$ is a parameter which controls the spread of the beam,
$\theta$ is the azimuthal angle of the momentum vector, and $p_{0}$ is the
magnitude of the momentum of the photon beam, which we arbitrarily set to unity.
We do not take into account a spread in the {\it magnitude} of the
momentum because the magnitude, $\omega$, is just a constant
multiplying the momentum 4-vector and so
\begin{eqnarray}\label{eq:magproof}
    \Lambda p^{\mu} & = & \Lambda ( \omega, \omega
    \hp ) \nonumber\\
    & = & \Lambda \omega ( 1, \hp ) \\
    & = & \omega \Lambda ( 1, \hp )\; . \nonumber
\end{eqnarray}
Inserting this result into Eq.~(\ref{epsilongauge}), we see that the
$\omega$-dependence cancels.

\begin{figure}[htb]
    \includegraphics[clip=true,width=\linewidth]{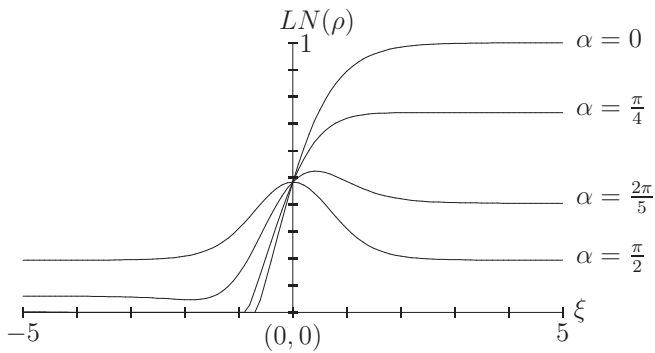}
    \caption{\label{fig:angle} Log negativity of the spin as a
    function of rapidity shown for various boost directions.
    $\alpha$ is the polar angle of the boost direction.  For
    all of the curves the angular spread is the same,
    $\sigma = 1.0$.}
\end{figure}

We now boost the state Eq.~(\ref{eq:hvstate}) and trace out the momentum degrees of
freedom to construct the polarization density matrix.  Because photons are spin-one
particles, they constitute three-level systems (even though they are constrained to
be transverse for any particular momentum). In order to calculate the entanglement
present in the quantum state, we therefore cannot use Wootters'
concurrence~\cite{Wootters98}, as it is only a measure of entanglement for
two-state quantum entangled systems. Instead, we use here ``log negativity", an
entanglement measure introduced by Vidal and Werner~\cite{vidal02}. This measure is
defined as
\begin{equation}\label{eq:negativity}
    LN(\rho) = \log_{2} \| \rho^{T_{A}} \| ,
\end{equation}
where $\| \rho \|$ is the trace norm and $\rho^{T_{A}}$ is the partial transpose of
$\rho$.  $LN(\rho)$ is a measure of the entanglement but is unable to
detect bound entanglement. We can now
calculate the change in log negativity explicitly for a Lorentz boost with rapidity
$\xi$ at an angle $\alpha$ with respect to the photon momentum, i.e.,
a Lorentz transformation
\begin{equation}\label{eq:boost}
    \Lambda = R_y(\alpha) L_z(\xi) R_y(\alpha)^{-1}\;,
\end{equation}
applied to Eq.~(\ref{eq:hvstate}). Fig.~\ref{fig:angle} summarizes the results of
varying the boost direction, $\alpha$, for a given spread, $\sigma_\theta$, and
shows that the entanglement can increase or decrease, depending on boost direction.
For $\alpha=0$, positive $\xi$ corresponds to boosting the photon in the direction
of the detector. Note that the entanglement at zero rapidity is only about half its
maximal value, because the angular spread of the momentum leaves the
spin degrees of freedom in a mixed state after tracing out momentum.
In general, boosts in the direction of motion tend to increase the entanglement to
saturation, while boosts away from it decrease it.  As $\alpha$ approaches
$\frac{\pi}{2}$, the effect on entanglement becomes symmetric.

\begin{figure}[htb]
    \includegraphics[clip=true,width=\linewidth]{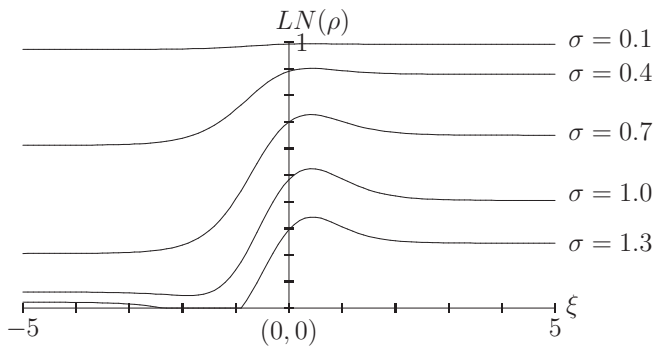}
    \caption{\label{fig:spread} Log negativity as a
    function of rapidity shown for beams of various angular
    spreads, $\sigma$.  For all of the above curves the boost
    direction $\alpha=\frac{2 \pi}{5}$.}
\end{figure}

Fig.~\ref{fig:spread} summarizes the effect of applying the boost in
Eq.~(\ref{eq:boost}) for varying spreads in the momentum distribution, for a
boost direction given by $\alpha = 2 \pi/5$.
Distributions with small spread, $\sigma_\theta \leq 0.1$, tend to change
entanglement only imperceptively, while for larger spread the entanglement changes
become more pronounced. Note that for $\sigma_\theta=1.3$ the entanglement becomes zero
(for boosts of negative rapidity) and then increases.  This appears to happen
because the momentum spread becomes so large that a significant portion of the beam
is in fact moving in the $-\hz$ direction. Because of the collimating effect that a
Lorentz boost has on the beam, the entanglement can actually increase in such
a situation.

We have derived the relativistic transformation law for photon polarizations, and
shown that the entanglement of polarization-entangled pairs of photon beams depends
on the reference frame. Boosting a detector (even at an angle) towards the beams
increases this entanglement because the momentum distribution is shrunk by the
boost (see also~\cite{PeresTerno02}). The type of entangled beams that we have
investigated in this letter are idealizations of realistic states that can be
created using parametric down-conversion. In principle, therefore, the effects
discussed here should become relevant as soon as linear-optics based quantum
technology is created that is placed on systems that move with respect to a
detector (or when the detector moves with respect to such a system).

We would like to thank Jonathan Dowling, and the members of the JPL
Quantum Computing Group, for useful discussions and encouragement. We
also acknowledge J\'anos Bergou for helpful suggestions. This work was
carried out at the Jet Propulsion Laboratory (California Institute of
Technology) under a contract with the National Aeronautics and Space
Administration, with support from the National Security Agency, the
Advanced Research and Development Activity, the Defense Advanced
Research Projects Agency, the National Reconnaissance Office, and the
Office of Naval Research.


\begin{thebibliography}{99}
\bibitem{dowling03}J.P. Dowling and G. J. Milburn, Phil. Trans. Roy. Soc. London
{\bf 361} (2003), to appear.
\bibitem{epr35}A. Einstein, B. Podolsky, and N. Rosen, Phys. Rev. {\bf 47},
777-780 (1935).
\bibitem{Bennettetal93}C.H. Bennett et al., Phys. Rev. Lett. {\bf 70},
1895(1993).
\bibitem{BennettWiesner92}C.H. Bennett and S.J. Wiesner, Phys. Rev. Lett. {\bf
69}, 2881 (1992).
\bibitem{dowling98}J.P. Dowling, Phys. Rev. A {\bf 57}, 4736 (1998).
\bibitem{NielsenChuang00}M.A. Nielsen and I.L. Chuang. \textit{Quantum Computation and
Quantum Communication }(Cambridge University Press, 2000).
\bibitem{Peresetal02}A. Peres, P.F. Scudo, and D.R. Terno, Phys. Rev. Lett. {\bf 88}, 230402 (2002).
\bibitem{Czachor97} M. Czachor,  Phys. Rev. A {\bf 55}, 72 (1997).
\bibitem{AlsingMilburn02}P.A. Alsing and G.J. Milburn, Quantum Inf. Comput. {\bf 2}, 487 (2002).
\bibitem{TerashimaUeda02}H. Terashima and M. Ueda, Int. J. Quant. Inf. {\bf 1} (2003).
\bibitem{GingrichAdami02} R.M. Gingrich and C. Adami, Phys. Rev. Lett. {\bf
    89}, 270402 (2002).
\bibitem{PeresTerno02} A. Peres and D.R. Terno, quant-ph/0208128.
\bibitem{PeresTerno03} A. Peres and D.R. Terno, Rev. Mod. Phys. {\bf 75} (2003), to
appear.
 \bibitem{wigner} E.P. Wigner, Ann. of Math. {\bf 40}, 149-204
    (1939).
\bibitem{tung} W.K. Tung, \textit{Group Theory in Physics} (World
    Scientific Publishing, 1985).
\bibitem{weinberg} S. Weinberg, \textit{The Quantum Theory of Fields},
    Vol. I (Cambridge University Press, 1995).
\bibitem{yndurain} F.J. Yndurain, \textit{Relativistic Quantum
    Mechanics and Introduction to Field Theory} (Springer Verlag,
    1996).
\bibitem{han} D. Han, Y.S.Kim and D. Son, Phys. Rev. D {\bf 31},
    328-330 (1985).
\bibitem{phinote}The choice of basis vectors $| h_{\bp} \rangle$ and
  $|v_{\bp} \rangle$ that guarantees polarization vectors as close as possible to
  the laboratory $\hx,\hy$ directions is obtained instead by using
  \begin{equation}
    \phi_{\bp} \rightarrow \psi_{\bp}
  \end{equation}
where $\psi_{\bp}$ is such that
\begin{equation}
\tan (2 \psi_{\bp}) = \tan (2
\phi) \frac{2 \cos(\theta)}{1 + \cos^2 (\theta)}\ .
\end{equation}
However, we used $\phi_{\bp}$ in the present calculations because it is
approximately equal to $\psi_{\bp}$ except when $\theta$ is large, and much more
convienient for our numerical simulations. We carried out a subset of the
calculations displayed in Figs. 2 and 3 using the optimal angle $\psi_{\bp}$, and
found essentially unchanged results.
\bibitem{Wootters98}W.K. Wootters, Phys. Rev. Lett. {\bf 80}, 2245 (1998).
\bibitem{vidal02} G. Vidal and R.F. Werner, Phys. Rev. A {\bf 65},
    032314 (2002).
\end{thebibliography}
\end{document}